\documentstyle[12pt]{article}
\begin{document}
\setlength{\baselineskip}{.7cm}
\renewcommand{\thefootnote}{\fnsymbol{footnote}}
\newcommand{\lp}{\left(}
\newcommand{\rp}{\right)}

\sloppy

\begin{center}
\centering{\Large \bf Stock market crashes, Precursors and Replicas}
\end{center}

\begin{center}
\centering{Didier Sornette $^{1,3}$, Anders Johansen $^{1,3}$ and Jean-Philippe
Bouchaud$^{2,3}$\\
{\it 1) Laboratoire de Physique de la Mati\`ere Condens\'ee,
CNRS URA 190\\ Universit\'e de Nice-Sophia Antipolis, B.P. 71\\ Parc
Valrose, 06108 Nice Cedex 2, France}

{\it 2) Service de Physique de l'Etat Condens{\'e}, CEA-Saclay\\
91191 Gif sur Yvette CEDEX, France}

{\it 3) Science \& Finance, 109-111 rue V. Hugo, 92 532 Levallois, France. }
}
\end{center}
\vskip .2in
\renewcommand{\thefootnote}{\fnsymbol{footnote}}

{\bf Abstract} \\
We present an analysis of the time behavior of the $S\&P500$ (Standard and
Poors) New York stock exchange index before and after the October 1987 market
crash and identify precursory patterns as well as aftershock signatures and
characteristic oscillations of relaxation. Combined, they all suggest a
picture of a kind of dynamical critical point, with characteristic log-periodic
signatures, similar to what has been found recently for earthquakes. These
observations are confirmed on other smaller crashes, and strengthen the view of
the stockmarket as an example of a self-organizing cooperative system.

\vskip 1cm
{\bf R\'esum\'e} \\
Nous pr\'esentons une analyse du comportement de l'indice boursier
americain $S\&P500$ avant et apr\`es le crash d'octobre 1987.
Nous identifions des motifs pr\'ecurseurs ainsi que
des oscillations de relaxation et des signatures de r\'epliques apr\`es le
crash. Ces caract\'eristiques sugg\`erent toutes ensembles que ce crash peut
\^etre vu comme une sorte de point critique dynamique, poss\'edant
des signatures sp\'ecifiques log-p\'eriodiques, comme on l'a
d\'ecouvert pr\'ec\'edemment pour les tremblements de terre. Ces observations
sont confirm\'es sur d'autres crashes plus petits et
renforcent le concept d'un march\'e mondial vu comme
un exemple de syst\`eme auto-organis\'e complexe.

\pagebreak

\section{The October 1987 crash}

{}From the opening on October 14, 1987 through the market close on October 19,
major indexes of market valuation in the United States declined by 30 percent
or more. Furthermore, all major world markets declined substantially in the
month, which is itself an exceptional fact that contrasts with the usual modest
correlations of returns across countries and the fact that
stock markets around the world are amazingly diverse in their organization
\cite{krach87}.

 In local currency units, the minimum
decline was in Austria ($-11.4\%$) and the maximum was in Hong Kong
($-45.8\%$). Out of $23$ major industrial countries \cite{countries}, $19$ had
a
decline greater than $20\%$. Contrary to a common belief, the US was not the
first to decline sharply. Non-Japanese Asian markets began a severe decline on
October $19$, 1987, their time, and this decline was echoed first on a number
of
European markets, then in North American, and finally in Japan. However, most
of
the same markets had experienced significant but less severe declines in
the latter part of the previous week. With the exception of the US and Canada,
other markets continued downward through the end of October, and some of these
declines were as large as the great crash on October $19$.

A lot of work has been carried out to unravel the origin(s) of the crash,
notably
in the properties of trading and the structure of markets; however, no
clear cause has been singled out. It is noteworthy that the strong market
decline during October 1987
followed what for many countries had been an unprecedented market increase
during the first nine months of the year and even before. In the US market for
instance, stock prices advanced $31.4 \%$ over those nine months. Some
commentators have suggested that the real cause of October's decline was that
over-inflated prices generated a speculative bubble during the earlier period.

However, the analysis on univariate associations and multiple
regressions of these various factors which have been carried out \cite{krach87}
 conclude that it is not clear at all what was the origin of the crash. The
most precise statement, albeit somewhat self-referencing, is that the most
statistically significant explanatory variable in the October crash can be
ascribed to the normal response of each country's stock market to a worldwide
market movement. A world market index was thus constructed \cite{krach87}
by equally weighting the local currency indexes
of $23$ major industrial countries \cite{countries} and normalized to $100$ on
september $30$. It fell to $73.6$ by October 30. The important result is that
it was found to be
statistically related to monthly returns in every country during the period
from the beginning of $1981$ until the month before the crash, albeit with a
wildly varying magnitude of the responses across countries \cite{krach87}.
This correlation was
found to swamp the influence of the institutional market characteristics.
This signals the possible existence of a subtle but nonetheless present
world-wide cooperativity.

In addition, we also note the highly nonlinear (threshold
like) behavior of traders, following positive and negative feedback
patterns \cite{Arthur}. This, in addition to these above facts and the
preliminary understanding of market self-organization provided by
simple statistical models \cite{Arthur}, lead us to
ask whether the October $1987$ crash could not be the result
of a worldwide cooperative phenomenon, with signatures in analogy with
critical phase transitions in physics. Here scale invariance and
self-similarity are the dominant concepts, which have proven
extremely useful in non-equilibrium driven systems, such as earthquakes,
avalanches, crack propagation, traffic flow to mention a few.

\section{Evidence for cooperative behavior and log-periodic oscillations}

\subsection{Precursory pattern}

Fig.1 shows the evolution as a function of time of the New York stock exchange
index $S\&P500$ from July $1985$ to the end of $1987$.
 The crosses represent the best fit to a constant
rate hypothesis corresponding to an average return of about $30 \%$ per year.
This first representation does not describe the apparent overall
acceleration before the crash, occurring already more than a year in advance.
This acceleration ({\it cusp}-like shape) is represented by
the monotonic line corresponding to a
fit of the data by an pure power law:
$$
\label{lp1} F_{pow}\lp t \rp  = A_1+B_1\lp t_c-t \rp ^{m_1}  , \eqno(1)
$$
where $t_c$ denotes the time at which the powerlaw fit of the $S\&P500$
presents a diverging slope, announcing an imminent crash. Since the ''noise''
content of $S\&P500$ is not known, a $\chi^2$-statistic cannot be calculated
in order to
qualify the fit. Instead, we have used the variance of the fit defined as
$\mbox{var}\lp f \rp = \frac{1}{N-n}\sum_{i=1}^N \lp y_i - f \lp t_i\rp \rp^2$,
where
$n$ is the number of free variables in $f$. (This assumes that the errors
are normally distributed, which is a reasonable null-hypothesis.) The ratio
of two variances corresponding to two different hypothesis is now the
qualifying statistic. For the constant rate hypothesis to that of the
power-law, we find a ratio $\mbox{var}_{exp}/\mbox{var}_{pow} \approx 1.1$
indicating a slightly better performance of the power law in capturing
the acceleration, the number of free variables being the same $(2)$.

However, already to the naked eye, the most striking feature in
this acceleration is the presence of systematic deviations. Inspired by the
analogy with critical phenomena, we have  fitted this structure by the
following mathematical expression
$$
\label{lp2} F_{lp} \lp \tau \rp  = A_2+B_2\lp \tau_c-\tau \rp ^{m_2} \left[
1+C\cos \lp
\omega \log \lp \tau_c-\tau \rp \rp \right] , \eqno(2)
$$
where $\tau = t/T$ is the time in units of $T$ and we use natural logarithm.
The
time scale $T$ comes about because the cosine is expected to have some phase
$\Psi$ defined by $\cos \lp \omega \log \lp t_c-t \rp - \psi\rp$. We can always
change variable with $\Psi =  \omega \log T$, which allows us to retrieve the
notation used in  Eq.(2). This shows that the phase $\Psi$ is therefore nothing
but a time scale. This equation is the first
Fourier component of a general log-periodic correction to a pure power law
for an observable exhibiting a cusp singularity at the time $t_c$ of
the crash, i.e. which becomes scale-invariant at the critical point
\cite{SS}.  Eq. $\lp 2\rp$ was first proposed to fit experimental measurements
of acoustic emissions prior to rupture of heterogeneous composite systems
stressed up to failure \cite{Anifrani}. It has also been observed to fit the
dependence of the released strain on the time to rupture for various large
Californian earthquakes and the seismic activity of the Aleutian-Island
seismic zone \cite{SS} as well as precursors to the recent Kobe earthquake
in Japan \cite{Johansen}. Beside, `complex exponents' (i.e. log-periodic
corrections to power laws)  have recently been found in a variety of
physical systems which constitute paradigms of self-organization and
complexity \cite{Salsor}. On a theoretical ground, they reflect the fact that
the system  is invariant under a {\em discrete} (rather than continuous) set
of dilatations only. While having been ignored for a long time,
it seems that complex exponents and their accompanying log-periodic
patterns are actually very common in Nature.

The Log-periodic corrections to scaling imply the existence of a hierarchy of
characteristic time intervals $t_c - t_n$, determined from the equation
$\omega \log (t_c-t_n) + T  =  n \pi$, which yields  $t_c - t_n =
\tau_0 \lambda^{n}$, with $\lambda = \exp{\pi \over \omega }$,
 $\tau_0 = \lambda^{-\frac{T}{\pi}}$ . For the October 1987 crash, we find
$\lambda \simeq 1.5-1.7$ (this value is remarkably universal and is
found approximately the same for other crashes and earthquakes) and $\tau_0
\simeq
0.85-0.95$ years. We expect a cut-off at short time scales (i.e. above $-n
\sim $ a few units) and also
at large time scales due to the existence of finite size effects. These
time scales $t_c - t_n$ are not universal but depend upon the specific market.
What is expected to be universal are the ratios $\frac{t_c - t_{n+1}}{t_c -
t_n}
= \lambda$. These time
scales could reflect the characteristic relaxation times associated with the
coupling between traders and the fundamentals of the economy.

The fit was performed as a minimization of the variance $\mbox{var}_{lp}$,
defined above, of the data. For the three linear variables $A_2$, $B_2$,
$C$, the minimization of the variance yields a set of three linear equations
which can be solved analytically, thus determining $A_2$, $B_2$ and $C$
as functions of the four nonlinear variables $m$, $t_c$, $\omega$ and
$T$. After this first step and replacing the analytical formulas of the
linear variables $A_2$, $B_2$ and $C$ in the expression of the variance, we
get a $4$-parameter fit where the remaining unknown variables are
$m$, $t_c$, $\omega$ and $T$. We claim this corresponds indeed to
a $4$-parameter fit (and not to a $7$-parameter fit) since we have used
an analytical constraint (here the minimization of the variance) to eliminate
$3$ unknown variables. This is completely similar, say, to the fit of a
probability distribution presenting a priori two unknown variables, the
normalizing factor $C$ and a characteristic decay rate $\mu$ ($C e^{-\mu x}$
for
an exponential distribution), in which the condition of normalization to $1$ of
the probability distribution imposes $C=\mu$ leading actually to a
$1$-parameter
fit. In addition, we checked that
the results are independent of the time unit used (which controls the $T$
variable). This was done by using either time measured in days
from the first point in the fit and also performing the fit with decimal
years counting from the turn of the century, giving exactly the same value
for $m,t_c,\omega$, implying that we face in fact an effective $3$-parameter
($m,t_c,\omega$) fit. Moreover, these three parameters $m$, $t_c$ and
$\omega$ are the most physically relevant, two of these ($m$ and $\omega$)
being
expected to exhibit some universality as discussed previously within the
renormalization group framework \cite{SS,Anifrani,Salsor}.

Due to the ``noisy'' nature of the data and the fact that we are performing
a minimization of the variance with respect to the four remaining
non linear parameters $m,t_c,\omega$ and $T$,
the $5$-dimensional space of the
$\mbox{var}_{lp}$  as a function of $m$, $t_c$, $\omega$ and $T$ has in
general several local minima. Hence, a preliminary restricted search (so-called
Taboo search \cite{taboo}) was performed before the full $4$-parameter fit was
executed, ensuring that the global minimum was found. This search was
done on a grid paving the two-dimensional space
($t_c$, $\omega$): for each given couple ($t_c$, $\omega$), we minimize the
variance with respect to the two other parameters and plot the resulting
variance as a function of $t_c$ and $\omega$. Finding the local minima of the
variance on this grid, we then launch a simplex algorithm on the four non
linear parameters $m$, $t_c$, $\omega$ and $T$.
The estimation of the
position of the critical time $t_c$ is found within a few days from the actual
crash time and the critical exponent $m$ is $m_2=0.33$.  The ratio between
$\mbox{var}_{lp}$ and that of the two other hypothesis is more than a factor of
$3$, which very clearly establishes $F_{lp}$ as the best performing fit among
the three proposed.

We also scanned regions without crashes to ascertain the absence of significant
log-periodic fluctuations there.

\subsection{Aftershock patterns}

If the concept of a crash as a kind of critical point has any value, we
should be able to identify post-crash signatures of the underlying
cooperativity. In fact, we should expect an at least qualitative symmetry
between patterns before and after the crash. In other words, we should be
able to document the existence of a critical exponent as well as
log-periodic oscillations on relevant quantities after the crash. We have
found such a signature in the variance (not to be confused with the variance
of the fit) of the $S\&P500$ index, implied from the $S\&P500$ options.

The term ''implied variance'' has the following
meaning. To understand what it means, one must first recall what is an option:
this financial instrument is nothing more than an insurance that can be bought
or sold on the market to insure oneself against unpleasant price variations
\cite{Bousor}.
The price of an option on the $S\&P500$ index is therefore a function
of the variance (so-called volatility) of the $S\&P500$. The more volatile, the
more fluctuating, the more risky is the $S\&P500$, the more expensive is the
option. In other words, the price of an option on the market
reflects the value of the variance of the stock as estimated by the market with
its offer-and-demand rules.
In practice, it is very difficult to have a good model for
market price volatilities or even to measure it reliably.
The standard procedure is then to see what the market forces decide for
the option price and then determine the implied volatility by inversion of the
Black and Scholes formula for option pricing \cite{RefGen,Bousor}.

Fig.2 presents the time evolution of the implied variance of the $S\&P500$
index after the crash, taken from \cite{finance}. As expected, the variance
decreases dramatically after the crash, while exhibiting
characterizing log-periodic oscillations.

Note the long time scale covering a period of the order of a year
involved in the relaxation of the volatility after the
crash to a level comparable to before the crash.
We also note that the $S\&P500$ index as well as others worldwide have remained
around the immediate of the crash level for a long time. For instance,
by February 29,1988, the world index stood at 72.7 (reference $100$ on
September 30, 1987). Thus, the price level established in the October crash
seems to have been a virtually unbiased estimate of the average price level
over
the subsequent months. Note also that the present value of the
$S\&P500$ index is much larger than it was even before the october 1987 crash,
showing again that nothing fundamental happened then.
All this is in support of the idea of a critical point,
according to which the event is an intrinsic signature of a
self-organization of the markets worldwide.

Our analysis with eq.(2), with
$t_c-t$ replaced by $t-t_c$ gives again an estimation
of the position of the critical time $t_c$, which is found within a few days.
The critical exponent is now $m_2 = -1.2$. The ratio of $\mbox{var}_{lp}$
to $\mbox{var}_{pow}$ and $\mbox{var}_{exp}$, respectively, is $\approx 2$,
the power law again performing slightly better than an exponential
relaxation hypothesis.

We have found another striking signature of the cooperative behavior of the
US market by analyzing the time evolution of the $S\&P500$ index over
a time window of a few weeks after the October 19 crash. A fit shown in
Fig.3 with a exponentially decaying sinusoidal function suggests that the US
market behaved, during a few weeks after the crash, as a {\it single}
dissipative harmonic oscillator. We think that this signature strengthens the
view of a market as a cooperative self-organizing system, presenting powerlaw
distributions, large events in possible coexistence with synchronized behavior.
Such properties have been indeed documented recently in models with
 threshold dynamics showing the generic
coexistence between critical self-organization and a large
'avalanche' regime corresponding to synchronization of threshold oscillators
\cite{synchro}. For the October $19$, 1987 crash, we find that the
characteristic decay time as well as the period of the oscillations are about
a week.

\section{Discussion}

We have found evidence of log-periodic structures in several others crashes
in a variety of markets \cite{longerwork}, paralleling previous similar
observations on earthquakes \cite{SS,Anifrani,Johansen}.
We suggest that this reflects the fundamental
cooperative nature of the behavior of stock markets. In general,
cooperative behaviors in complex systems cannot be
reduced to a simple decomposition on elementary causes, in agreement with the
observation  \cite{krach87} that no single source \cite{cause} has been
identified as a key factor in the October 1987 crash. One must rather look
from a more global view point in which the crash can emerge ''naturally''
as an intrinsic signature of the functioning of the market.

To rationalize these observations, we will report elsewhere \cite{longerwork}
on a simple
model of stockmarket speculation leading to a crash based on the
existence of positive feedback interactions in which traders exchange
information according to a hierarchical structure. This structure
is intended  to model the organization of the market in
the world, where at the highest level of the hierarchy, we find the
``currency and trading-blocks'' (Yen, US\$, D-mark, ...), at the level
immediately below we  have countries, at the level below the major banks and
institutions within a country, at the level below the various departments of
the
banks, etc. Hierarchy or, what is the same, discrete scale invariance, be it
structurally built-in or dynamically generated, has been recognized as the key
ingredient to obtain log-periodic behavior \cite{SS,Salsor}. As expected, the
solution of the model indeed shows the existence of a critical point which
can be
identified as
the crash and of well-defined precursory and aftershock log-periodic patterns.
Although the model is rather ad-hoc, these results
make more plausible our above observation of a qualitative symmetry
in the critical behavior of the market before and after the crash. This model
analyzes a situation of pure speculation, based on the tendency for traders
to imitate each others. When a series of buy orders, say, are issued,
an acceleration of demand results, which is self-strengthening. This
acceleration
cannot be sustained indefinitely and, at some threshold, a crash ends
this sequence.

To sum up, the acceleration described by a power law is the signature of
a critical point. The log-periodic oscillations are the signature of
discrete scale invariance in the trading structure given above. There are
several mechanisms that can
generate this remarkable structure; for instance a built-in hierarchical
structure or irreversible non-linear intermittent
dynamics are know to generate these patterns \cite{SS,Salsor}.

It is intriguing that the log-periodic structures documented here
bear some similarity with the `Elliott waves' of technical analysis
\cite{Elliott}. Technical analysis in finance can be broadly defined as the
study of financial markets, mainly using graphs of stock prices as a function
of time, in the goal of predicting future trends. A lot of efforts has been
developed in finance both by academic and trading institutions and more
recently by physicists (using some of their statistical tools developed to deal
with complex times series) to analyse past data to get informations on
the future. The 'Elliott wave' technique is probably the most famous in this
field. It has been introduced in the $1930$'s, based on observations on the
human
(trader) psychology on one hand and from analogies with the mathematical theory
of numbers and more precisely the theory of Fibonacci numbers on the other
hand.
It describes the time series of a stock price as made of different ''waves''.
These different waves are in relation with each others through the Fibonacci
series $F_{n+2} = F_{n+1} + F_n$ (with $F_0 = F_1 = 1$). It is easy to
show that ${F_{n+1} \over F_n}$ converges to a constant (the so-called golden
mean $g \simeq 1.618$), implying an approximate geometrical series of time
scales $F_{n+1} \simeq g F_n$ in the underlying waves, compatible with our
above
estimate for the ratio $\lambda \simeq 1.5-1.7$. We speculate that the
`Elliott waves', so strongly rooted in the financial analysts' folklore,
could be a signature of an underlying critical structure
of the stockmarket.

Acknowledgements : We thank O. Couvreur and J.-P. Aguilar for stimulating
discussions and J.-C. Roustan for help in the analysis of figure 2.

After completion of this work, we learned that James A. Feigenbaum and Peter
G.O. Freund (cond-mat/9509033) have obtained, independently,
very similar results to ours.

\pagebreak

\pagebreak

{\Large \bf Figure captions}

\begin{itemize}

\item Fig.1 : evolution as a function of time of the New York stock exchange
index $S\&P500$ from July $1985$ to the end of $1987$ ($557$ trading days).
The $+$ represent a constant return increase of $\approx 30 \%$/year and
had $var\lp F_{exp} \rp \approx 113$.
The best fit to a power-law gives $A_1 \approx 327$, $B_1  \approx -79$,
$t_c  \approx 87.65$, $m_1  \approx 0.7$ and $\mbox{var}_{pow} \approx 107$.
The best fit to eq.(2) gives $A_2 \approx 412$, $B_2 \approx -165$,
$t_c \approx 87.74$, $C \approx 12$, $\omega \approx 7.4$, $T =2.0$,
$m_2 \approx 0.33$ and $\mbox{var}_{lp} \approx 36$. One can observe
four well-defined oscillations fitted by eq.(2), before finite size effects
limit the theoretical divergence of the acceleration, at which point the
bubble ends in the crash. All the fits are carried over the whole time interval
shown, up to $87.6$. The fit with eq.(2) turns out to be very robust with
respect
to this upper bound which can be varied significantly.

\item Fig.2 : Time evolution of the implied variance in $\log$-scale of the
$S\&P500$ index after the crash, taken from \cite{finance}. The $+$ represent
an
exponential decrease with $var\lp F_{exp} \rp \approx 15$.
The best fit to a power-law, represented by the monotonic line, gives $A_1
\approx 3.9$, $B_1 \approx 0.6$,  $t_c = 87.75$, $m_1 \approx 1.5 $ and
$\mbox{var}_{pow} \approx 12$. The best fit to eq.(2) with $t_c - t$ replaced
by
$t-t_c$ gives  $A_2 \approx 3.4$, $B_2 \approx 0.9$, $t_c \approx 87.77$, $C
\approx 0.3$,  $\omega \approx 11$, $m_2 \approx -1.2$ and $\mbox{var}_{lp}
\approx 7$. One can observe six well-defined oscillations fitted by eq.(2).

\item Fig.3 : Time evolution of the $S\&P500$ index over
a time window of a few weeks after the October 19 crash. The fit
with an exponentially decaying sinusoidal function suggests that a good
model for the short-time response of the US market is a {\it single}
dissipative harmonic oscillator.

\end{itemize}


\begin{thebibliography}{99}
\bibitem{cause} Among the features that have been suggested to be relevant are
the computerized trading, the auction system itself, the presence or absence
of limits on price movements, regulated margin requirements, off-market and
off-hours trading (continuous auction and automated quotations), the presence
or
absence of floor brokers who conduct trades but are not permitted to invest on
their own account, the extend of trading in the cash market versus the forward
market, futures and options trading, the identity of traders (i.e. institutions
such as banks or specialized trading firms), the significance of transaction
taxes.
\end{thebibliography}
\end{document}